\documentclass[preprint2]{aastex}

\shorttitle{RHESSI CDS}

\shortauthors{Milligan et al.}

\begin{document}

\title{Observational Evidence for Gentle Chromospheric Evaporation During the Impulsive Phase of a Solar Flare}

\notetoeditor{the contact email is r.milligan@qub.ac.uk and is the only one which should appear on the journal version}

\author{Ryan O. Milligan\altaffilmark{1,3}, 
        Peter T. Gallagher\altaffilmark{2,3,4},
        Mihalis Mathioudakis\altaffilmark{1}, and
        Francis P. Keenan\altaffilmark{1}}

\altaffiltext{1} {Department of Physics and Astronomy, Queen's University Belfast, Belfast, BT7 1NN, Northern Ireland.}
\altaffiltext{2} {School of Physics, Trinity College Dublin, Dublin 2, Ireland.}
\altaffiltext{3} {Laboratory for Solar and Space Physics, NASA Goddard Space Flight Center, Greenbelt, MD 20771, U.S.A.}
\altaffiltext{4} {L-3 Communications GSI.}

\begin{abstract}  

Observational evidence for gentle chromospheric evaporation during the
impulsive phase of a C9.1 solar flare is presented using data from the {\it
Reuven Ramaty High-Energy Solar Spectroscopic Imager} and the Coronal
Diagnostic Spectrometer on board the {\it Solar and Heliospheric Observatory}.
Until now, evidence for gentle evaporation has often been reported during the
decay phase of solar flares, where thermal conduction is thought to be the
driving mechanism. Here we show that the chromospheric response to a low flux
of nonthermal electrons ($\geq$5$\times$10$^{9}$~ergs~cm$^{-2}$~s$^{-1}$)
results in plasma upflows of 13$\pm$16, 16$\pm$18, and 110$\pm$58~km~s$^{-1}$
in the cool \ion{He}{1} and \ion{O}{5} emission lines and the 8~MK \ion{Fe}{19}
line. These findings, in conjunction with other recently reported work, now
confirm that the dynamic response of the solar atmosphere is sensitively
dependent on the flux of incident electrons.

\end{abstract}


\section{INTRODUCTION} 
\label{intro} 

During the impulsive phase of a solar flare, accelerated electrons propogate
along closed magnetic field lines to the dense, underlying chromosphere, where
they lose their energy via Coloumb collisions and heat the local plasma. The
resulting expansion of this plasma is known as ``chromospheric evaporation''.
From the hydrodynamic simulations of \cite{fish85a,fish85b,fish85c}, and more
recently \cite{abbe99}, the solar atmosphere is predicted to respond in one of
two ways, depending on the flux of accelerated nonthermal electrons. 

\begin{figure*}[!t]
\begin{center}
\includegraphics[height=16.5cm,angle=90]{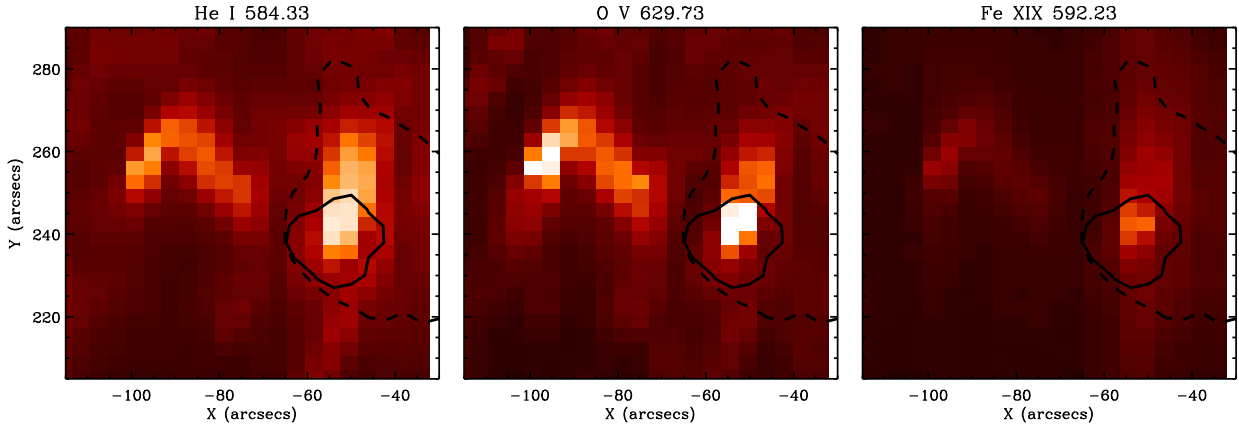}

\caption{CDS images obtained in the \ion{He}{1} (log T = 4.5), \ion{O}{5} (log T
= 5.4), and \ion{Fe}{19} (log T = 6.9) emission lines observed during the
impulsive phase of the 2002 July 15 solar flare. {\it RHESSI} 6--16~keV ({\it
dotted line}) and 16--50~keV ({\it solid lines}) contours are overlaid, drawn at
5\% and 10\% of the peak intensity, respectively.}

\label{hsi_cds}
\end{center}
\end{figure*}

For electron fluxes $\leq$10$^{10}$~ergs~cm$^{-2}$~s$^{-1}$, the evaporated 
plasma flows upwards at several tens of kilometers per second, with no
associated downflows. This process is termed `gentle' evaporation. Gentle
evaporation may also occur during the decay phase, when the upflows are driven
by thermal conduction rather than electron beam heating (\citealt{anti78}).
Many studies have reported evidence for conduction driven evaporation during
the decay phase of solar flares (\citealt{schm87}; \citealt{zarr88};
\citealt{czay01}; \citealt{berl05}). In each of these studies it was concluded
that the late phase evaporation was caused by heat conduction along field lines
connecting the chromosphere to the corona. Evidence for gentle evaporation was
also presented by \cite{bros04} during flare precursor events, but the
mechanism responsible could not be verified. To date, there has been no direct
evidence for gentle evaporation due to nonthermal electrons during the
impulsive phase of a solar flare.

At high nonthermal electron fluxes
($\gtrsim$3$\times$10$^{10}$~ergs~cm$^{-2}$~s$^{-1}$), the chromosphere is
unable to radiate at a sufficient rate and consequently expands rapidly. The
heated chromospheric plasma ($\sim$10$^{7}$~K) expands upwards at hundreds of
kilometers per second in a process known as `explosive' evaporation. The
overpressure of the flare plasma relative to the underlying chromosphere also
causes cooler, more dense material to expand downwards at tens of kilometers per
second. This process is known as `chromospheric condensation'. A strong case for
explosive evaporation was presented by \cite{bros04}, who reported oppositely
directed flows using the Coronal Diagnostic Spectrometer (CDS; \citealt{harr95})
on board the {\it Solar and Heliospheric Observatory (SOHO)} during a hard X-ray
(HXR) burst. A more recent study by Milligan et al. (2006; hereafter referred to
as Paper I) also found these flows patterns, but critically, were able to derive
the properties of the driving electron beam using simultaneous HXR imaging and
spectroscopy from the {\it Reuven Ramaty High-Energy Solar Spectroscopic Imager}
({\it RHESSI}; \citealt{lin02}).

In this {\it Letter} we present the first observational evidence for gentle
chromospheric evaporation due to nonthermal electrons during the impulsive phase
of a solar flare.  A brief overview of the instruments and data analysis is
given in \S~\ref{obs} (a more detailed description can be found in Paper I). Our
results are then presented in \S~\ref{results}, while a discussion and
conclusions are given in \S~\ref{disc}. 

\section{OBSERVATIONS}
\label{obs}

Our study focuses on a compact GOES C9.1 flare, which began at 11:40:08~UT on
2002 July 15. The flare occured close to the solar meridian (-40$\arcsec$,
232$\arcsec$) during a joint observing plan between {\it RHESSI} and other
ground- and space-based observatories. Unfortunately, there were no
complimentary EUV images available from either the EUV Imaging Telescope (EIT)
or the {\it Transition Region and Coronal Explorer} during the event. 

The CDS observations reported here were aquired with the {\it FLARE\_AR}
observing sequence (see Paper I for details). Images taken during the impulsive
phase in the \ion{He}{1}  (584.33~\AA), \ion{O}{5} (629.73~\AA), and
\ion{Fe}{19} (592.23~\AA) emission lines are shown in Figure~\ref{hsi_cds}.
Spectra from each CDS pixel were fitted with a broadened Gaussian profile
\citep{thom99}, for each of the spectral windows. Velocities were found by
measuring Doppler shifts relative to quiet-Sun spectra, which were assumed to be
emitted by stationary plasma. A heliographic correction was also applied
assuming purely radial flows.

\begin{figure}[!t]
\begin{center}
\includegraphics[width=8.0cm]{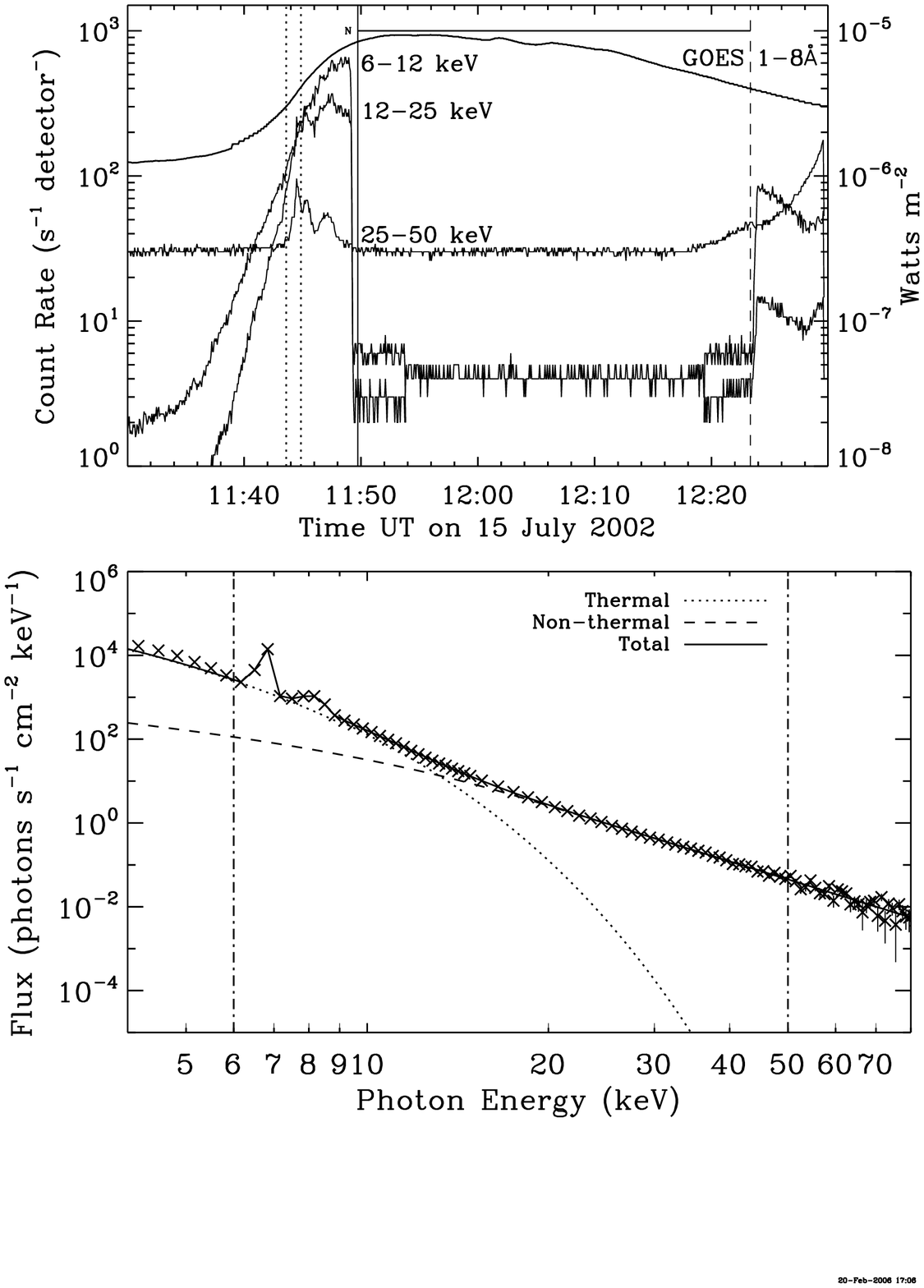}

\caption{{\it Top panel}: {\it RHESSI} observing summary data (corrected for
changes in attenuator states) from the 6--12, 12--25, and 25--50~keV bands. The
vertical dotted lines indicate the time interval over which images and spectra
were obtained, corresponding to the observation of significant upflows with CDS.
The bar denoted `N' at the top of the plot illustrates the period that {\it
RHESSI} was in eclipse. Overplotted is the {\it GOES} 1-8~\AA~curve. {\it
Bottom panel}: Portion of the {\it RHESSI} spectrum integrated over the time
range given above. The energy range 6--50~keV lying between the vertical
dot-dashed lines was fitted assuming an isothermal component ({\it dotted
curve}) and a thick-target bremsstrahlung component ({\it dashed curve}).}

\label{hsi_ltc_spec}
\end{center}
\end{figure}

{\it RHESSI} is an imaging spectrometer capable of observing X- and $\gamma$-ray
emission over a wide range of energies ($\sim$3~keV--17~MeV). The thin
attenuators on {\it RHESSI} were in place during part of the impulsive phase of
this event, thus limiting the energy range to $\gtrsim$6~keV. Flare emission was
not observed above $\sim$50~keV. The observing summary data for the energy
ranges 6--12, 12--25, and 25--50~keV, are shown in the top panel of
Figure~\ref{hsi_ltc_spec}. As {\it RHESSI} went into eclipse at
$\sim$11:49:00~UT, the GOES lightcurve for this event has been overplotted for
completeness. Both the {\it RHESSI} images and spectra were obtained over a 76
second period from 11:43:36--11:44:52~UT to coincide with the time range over
which CDS observed blueshifts in the \ion{Fe}{19} line. This time interval is
indicated by two vertical dotted lines in the top panel of
Figure~\ref{hsi_ltc_spec} and includes the impulsive 25--50~keV HXR burst. {\it
RHESSI} images in two energy bands (6--16 and 16--50~keV) were reconstructed
using the {\it Pixon} algorithm \citep{hurf02}. Contours at 5\% and 10\% of the
peak intensity, respectively, in each band are overlayed on each EUV image in
Figure~\ref{hsi_cds}. The {\it RHESSI} spectrum was fitted assuming an
isothermal distribution at low energies and thick-target emission at higher
energies (bottom panel of Figure~\ref{hsi_ltc_spec}).

\section{RESULTS}
\label{results}

The thick-target model solution consistant with the {\it RHESSI} photon  
spectrum produced an electron distribution with a low-energy cutoff 
($\epsilon_c$) of $\sim$~20~keV, and a power-law index ($\delta$) of $\sim$~5.2;
a break energy of $\sim$20~keV in the electron spectrum corresponds to a break
energy of $\sim$16~keV for the associated photon spectrum. From the properties
of the inferred electron spectrum, the total power of nonthermal electrons was
found to be $\sim$8$\times$10$^{27}$~ergs~s$^{-1}$. The reconstructed 16--50~keV
image yielded an upper limit to the HXR source size of
$\sim$1.8$\times$10$^{18}$~cm$^{2}$, and the resulting flux of nonthermal
electrons was therefore found to be
$\geq$5$\times$10$^{9}$~ergs~cm$^{-2}$~s$^{-1}$.

\begin{figure*}[!t] 
\begin{center} 
\includegraphics[height=16.5cm,angle=90]{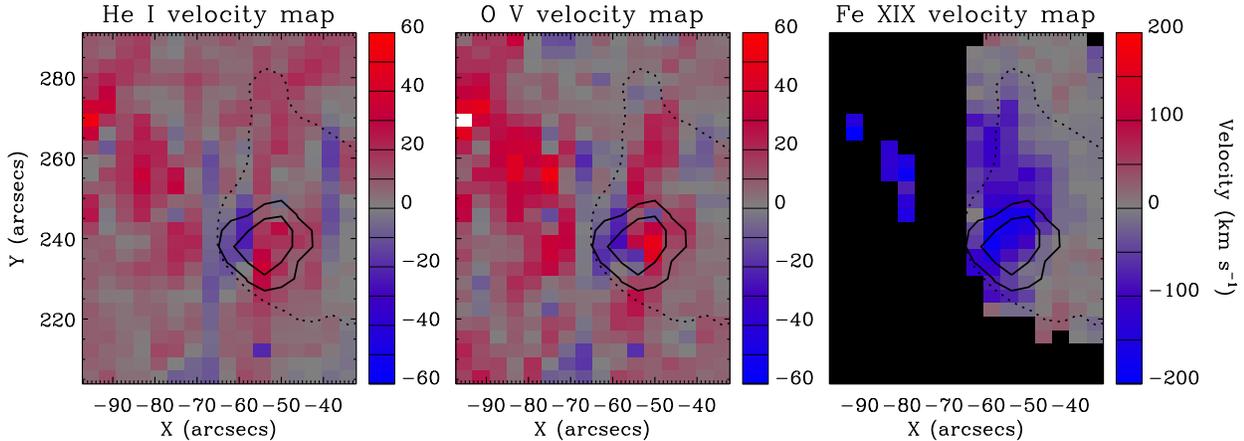}

\caption{Velocity maps in the \ion{He}{1}, \ion{O}{5}, and \ion{Fe}{19} lines.
Downflows are indicated by red pixels, while upflows are given in blue pixels.
The solid line denotes the {\it RHESSI} 16--50~keV contours at 10\% and 50\% of
the peak intensity, while the dotted contour shows the 6--16~keV emission at 5\%
of the peak intensity.}

\label{vel_maps} 
\end{center}
\end{figure*}

\begin{figure}[!t]
\begin{center}
\includegraphics[height=8.0cm,angle=90]{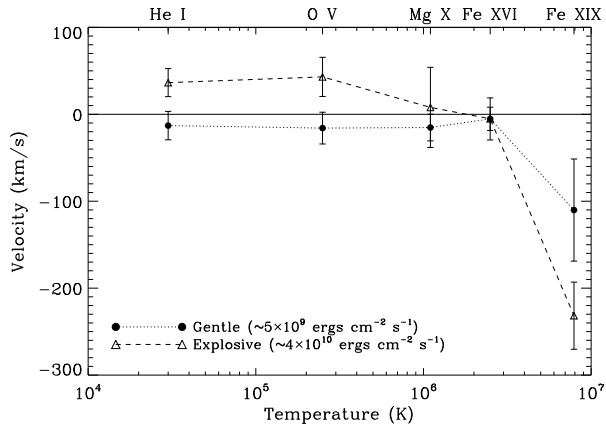}

\caption{Plasma velocity as a function of temperature for each of the five lines
observed using CDS. Positive velocities indicate downflows, while negative
values indicate upflows. The data points plotted with solid circles denote the
values presented in this study, while the triangles illustrate those values
presented in Paper I for the case of explosive evaporation. The dotted and
dashed lines connecting the points are added to guide the reader.}

\label{vel_plot}
\end{center}
\end{figure}

Figure~\ref{vel_maps} shows the spatial distribution of the plasma flows seen in
each of the \ion{He}{1}, \ion{O}{5}, and \ion{Fe}{19} lines. Net upflows of
13$\pm$16 and 16$\pm$18~km~s$^{-1}$ were observed in the \ion{He}{1} and
\ion{O}{5} maps by averaging over all CDS pixels within the 16-50~keV 10\%
contour as observed by {\it RHESSI}. Moderately strong upflows of
110$\pm$58~km~s$^{-1}$ were observed in the \ion{Fe}{19} map by averaging over
the same area. Weaker flows of $\lesssim$50~km~s$^{-1}$ were measured away from
this area and appear to be aligned with the thermal (6--16~keV) emission
observed by {\it RHESSI}. The mean velocity (from within the HXR emitting area)
as a function of temperature for each of the five emission lines observed by CDS
is shown in Figure~\ref{vel_plot}. Error bars represent a 1~$\sigma$
uncertainty. The values reported in Paper I for the case of explosive
evaporation (electron flux $\geq$4$\times$10$^{10}$~ergs~cm$^{-2}$~s$^{-1}$) are
overplotted.

\section{DISCUSSION AND CONCLUSIONS}
\label{disc}

Simultaneous X-ray and EUV observations of gentle chromospheric evaporation
during the impulsive phase of a C9.1 solar flare are presented using data from
{\it RHESSI} and {\it SOHO}/CDS. Until now, studies that reported evidence for
gentle evaporation were entirely focused on low-velocity
($\lesssim$~80~km~s$^{-1}$) mass motions observed during the decay phase of
solar flares (\citealt{schm87}; \citealt{zarr88}; \citealt{czay01};
\citealt{berl05}). In each of these cases, the mechanism responsible was
believed to be thermal conduction caused by the steep temperature gradients
between the high-temperature flare plasma and the cool, underlying chromosphere
\citep{anti78}. This was also motiviated by the fact that power-law spectra are
not observed during the decay phase of the majority of solar flares.

In this {\it Letter}, we report the first observational evidence for gentle
evaporation due to nonthermal electrons during the impulsive phase of a solar
flare as predicted by current theoretical models. Upflows of 13$\pm$16,
16$\pm$17, and 110$\pm$58~km~s$^{-1}$ as seen in the \ion{He}{1}, \ion{O}{5},
and \ion{Fe}{19} lines, respectively, result from a beam of non-thermal
electrons with a flux value of $\geq$5$\times$10$^{9}$~ergs~cm$^{-2}$~s$^{-1}$.
From Paper I, we reported that an order of magnitude higher flux
($\geq$4$\times$10$^{10}$~ergs~cm$^{-2}$~s$^{-1}$) gives rise to downflows of
$\sim$35 and $\sim$45~km~s$^{-1}$ in the \ion{He}{1} and \ion{O}{5} lines, and
upflows of $\sim$270~km~s$^{-1}$ in the \ion{Fe}{19} line. As a consequence, the
findings reported here and in Paper I support the theoretical models which
predict that the response of the solar chromosphere is sensitively dependent on
the flux of accelerated nonthermal electrons. Our results also support the
prediction that there exists a threshold value for the nonthermal electron flux
above which the chromosphere cannot efficiently radiate the deposited energy.
\cite{fish85a} proposed that this threshold value is
$\sim$3$\times$10$^{10}$~ergs~cm$^{-2}$~s$^{-1}$. Above this value, electron
fluxes were shown to begin to drive chromospheric condensation, a process
resulting from the high pressures reached by the rapidly expanding evaporated
material. The combination of the results presented here and in Paper I also lend
very strong evidence to support this principle. 

The recent hydrodynamic simulations of \cite{abbe99} and \cite{allr05} have been
developed to include more realistic electron beam parameters and a non-LTE
treatment of the solar atmosphere. These more detailed calculations still
provide the distinction between gentle and explosive evaporation for differing
nonthermal electron fluxes. These models will be further developed to include
higher temperature plasmas that will be observed by the EUV Imaging Spectrometer
(EIS) on board {\it Solar-B}, due for launch in late 2006. By combining EIS
observations with {\it RHESSI} data in the future, an even greater understanding
on the behaviour of high-temperature plasmas during solar flares will be
achieved. 

\acknowledgments   

This work has been supported by a Department of Employment and Learning
studentship in conjunction with a Cooperative Award in Science and Technology
from NASA Goddard Space Flight Center. F. P. K. is grateful to A. W. E.
Aldermaston for the award of a William Penny Fellowship. We would like to thank
Brian Dennis and the {\it RHESSI} team, Joe Gurman and Dominic Zarro at Goddard
for their continued support. {\it SOHO} is project of international
collaboration between the European Space Agency and NASA.

\end{document}